\documentclass[conference,letterpaper]{IEEEtran}
\usepackage[letterpaper, left=0.7in, right=0.65in, bottom=0.98in, top=0.75in]{geometry}

\usepackage{multicol}
\usepackage{cite}
\usepackage{multirow}
\usepackage{graphicx,color}
\usepackage{amssymb,amsmath,times,wasysym,amsthm,mathtools} 
\usepackage{srcltx,verbatim} 
\usepackage{epsfig}
\usepackage{breqn}
\usepackage{xcolor}
\usepackage[linesnumbered,ruled,vlined]{algorithm2e}

\title{Deep Learning-Based Device-Free Localization in Wireless Sensor Networks\vspace{-3mm}}

\author{\IEEEauthorblockN{Osamah A. Abdullah\IEEEauthorrefmark{1},
		Hayder Al-Hraishawi\IEEEauthorrefmark{2},
		and Symeon Chatzinotas\IEEEauthorrefmark{2}\\ 
				\IEEEauthorrefmark{1}\small Dept. of Electrical and Computer Engineering, Alma'moon University College, Baghdad, Iraq}
		\IEEEauthorrefmark{2}\small Interdisciplinary Centre for Security, Reliability and Trust (SnT), University of Luxembourg, L-1855, Luxembourg\\
		Email:  osamah.abdullah@wmich.edu, hayder.al-hraishawi@uni.lu, and symeon.chatzinotas@uni.lu
		\vspace{-5mm}}

\begin{document}
	\maketitle
	\thispagestyle{plain}
    \pagestyle{plain}

\begin{abstract}
Location-based services (LBS) are witnessing a rise in popularity owing to their key features of delivering powerful and personalized digital experiences. The recent developments in wireless sensing techniques make the realization of device-free localization (DFL) feasible in wireless sensor networks. The DFL is an emerging technology that utilizes radio signal information for detecting and positioning a passive target while the target is not equipped with a wireless device. However, determining the characteristics of the massive raw signals and extracting meaningful discriminative features relevant to the localization are highly intricate tasks. Thus, deep learning (DL) techniques can be utilized to address the DFL problem due to their unprecedented performance gains in many practical problems. In this direction, we propose a DFL framework consists of multiple convolutional neural network (CNN) layers along with autoencoders based on the restricted Boltzmann machines (RBM) to construct a convolutional deep belief network (CDBN) for features recognition and extracting. Each layer has stochastic pooling to sample down the feature map and reduced the dimensions of the required data for precise localization. The proposed framework is validated using  real experimental dataset. The results show that our algorithm can achieve a high accuracy of 98\% with reduced data dimensions and low signal-to-noise ratios (SNRs). 
\end{abstract} 
	

\section{Introduction}\label{sec:intro}
Contextualization and personalized services are of the most feature applications envisioned in the forthcoming sixth-generation (6G) wireless systems, and communication perception has been steadily growing  to become  one of the essential business application scenarios \cite{Zhou2022}. Therefore, it is particularly important to enhance the location-based services (LBS) to provide context-aware and personalized solutions for mobile users in diverse applications as well as for some critical scenarios such as patient tracking, intruder detection, and elderly surveillance \cite{kuutti2018}. Additionally, the precise localization of devices is fundamental in the context of Internet of Things (IoT) applications and future Industry 4.0 technologies. 
Further, the localization of smartphones has been used for contact tracing during the fight against the Covid-19 pandemic \cite{tomasin2021}. 
However, traditional localization techniques such as radio frequency identification (RFID) and global positioning system (GPS) are not always applicable because they need a tracking device to be attached to the target. To solve this issue, the concept of a passive wireless monitoring system, also known as device free localization (DFL), is introduced based on the wireless sensor networks \cite{Zhao2019,Alam2021}.

In DFL systems, wireless sensor nodes, also known as anchor points (APs), are used to sense and accumulate target positions via collaborative communications with other neighboring APs. Specifically, targets existence, absence, or appearing at different locations will definitely change the transmitting-receiving correspondences among the APs.
Accordingly, the position of the target can by analyzed/determined through the variations in the radio frequency (RF) signals. For instance, human movements have different effects on the received signal strength (RSS), which can be utilized to locate the target \cite{Kianoush2017}. In this setting, when a target enters the DFL system’s tracking area, certain RSS metrics are obtained and analyzed to estimate the target location. Namely, the tracking area is discretized into grids and the variations in RSS measurements due to the target movements can be regarded as a class of possibilities inside each grid cell, and hence, the DFL can be viewed as a classification problem.
This observation has motivated the researchers to utilize deep learning (DL) techniques to improve DFL performance \cite{Wang2017a}.

RSS signals in the wireless sensor networks are easily affected by the surrounding environment and the target movements, which add random noise to the collected RSS measurements and make them fluctuating as multimodal signals. Additionally, indoor localization is more challenging than outdoor localization due to the multi-path interference, scattering and reflecting from different sources such as walking people, furniture, walls, and other objects \cite{Feng2021}. Therefore, attaining a satisfactory localization accuracy requires deploying an adequate number of APs and collect large amounts of RSS measurements, which impose heavy  computational burdens to the DFL process, especially in the emergency scenarios when the positioning speed is necessary.  
In this context, various prior studies have considered machine learn (ML) approaches and the most popular DL algorithms such as K-nearest-neighbor (KNN) and support vector machines (SVM) to tackle the DFL problem \cite{Hong2015}.

Although conventional DL algorithms are effective in solving the classification problems, their performance within DFL systems diminishes with the high-dimensional data and occasionally fail to provide an accurate localization in a timely fashion \cite{Huang2018}. Inspired by this observation, this paper aims at developing a DFL framework that is able to efficiently process high dimensional data and capable of obtaining high levels of location accuracy in noisy environments. In this direction, a combination of convolutional deep belief network (CDBN) with feature autoencoding algorithm is applied for layer-wise learning. 
Specifically, CDBN is a hierarchical generative model that is used with probabilistic max-pooling to extract high-level features by replacing the restricted Boltzmann machine (RBM) in the deep belief networks that are used for the individual layers by convolutional RBMs, which has been proved as a useful algorithm for dimensionality reduction, collaborative filtering, feature learning and classification\cite{Lee2009}. 
In this framework, pooling layers are added between the convolutional layers and the training of the network is divided into two successive stages in order to reduce the dimension of upper layers. Specifically, first the pre-training stage where CDBN is used as a generative model to extract the features automatically from the RSS dataset, then the training stage is conducted by fine-tuning the parameters through the autoencoder, where the contrastive divergence is employed.
Finally, the learned features are merged into the autoencoder of the neural network to locate the target by using the training parameters.
In brief, our main contributions in this work can be summarized as follows:
\begin{itemize}
\item Applying the CDBN as a dimensionality reduction approach to solve the DFL classification problem by extracting high-level features from the wireless signals.
\item Developing a DL-based DFL framework that is efficiently able to   . The obtained results show that the proposed algorithm outperforms the conventional autoencoders.
\end{itemize}

The remainder of the paper is organized as follows.
The system model with the DFL problem formulation and the proposed approach are discussed in Section \ref{sec:system_model}. The simulation results and conclusion remarks are given in Section III and IV, respectively.

\begin{figure}[!t]
	\includegraphics[width=0.45\textwidth]{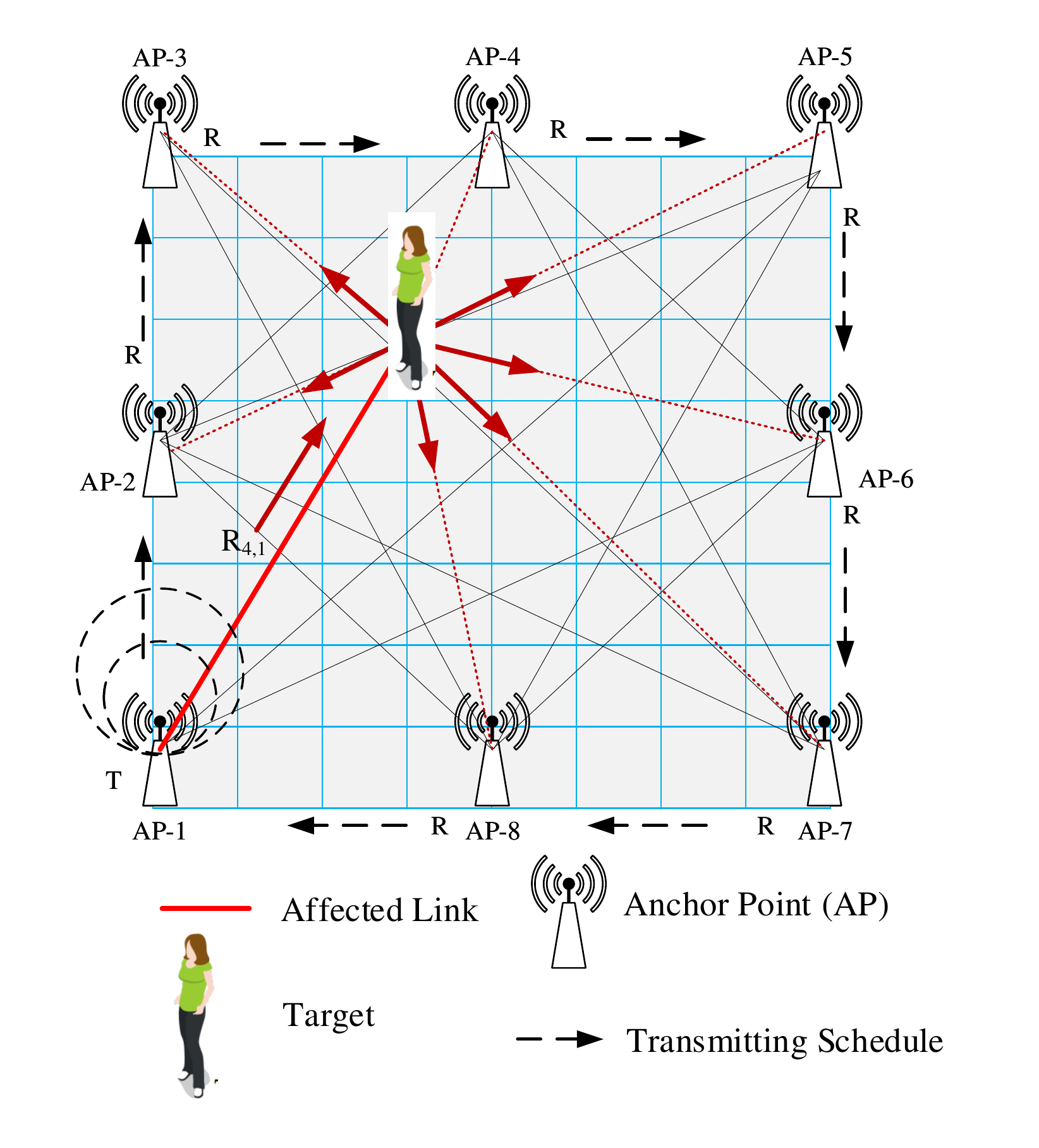} \vspace{-7mm}
	\caption{A schematic diagram  of a device-free localization system.} \label{fig:Fi1}\vspace{-4mm}
\end{figure}

\begin{figure*}[!t]
	\centering
	\setlength\fboxsep{0pt}
	\setlength\fboxrule{0.25pt}
	\includegraphics[width=0.85\textwidth]{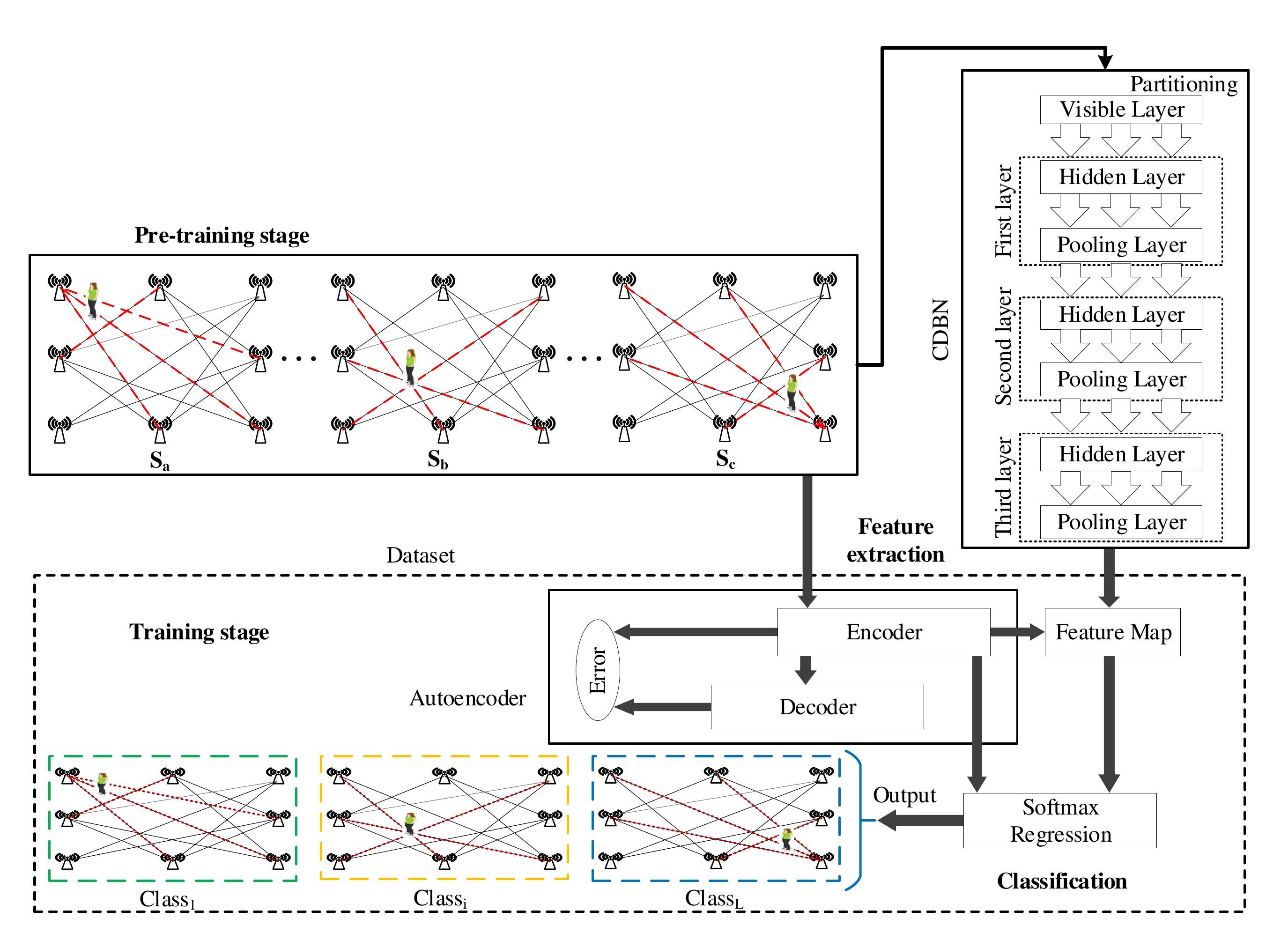} \vspace{-6mm}
	\caption{Illustration of the proposed framework for device free localization (DFL) using convolutional deep belief networks (CDBN).}\vspace{-4mm}
	\label{fig:Figure_2}
\end{figure*}

\section{DFL Problem Formulation}\label{sec:system_model}
We consider a DFL system consists of $N$ APs that are covering the area of interest $D$, and these APs are communicating with each other through wireless links, as shown in Fig. \ref{fig:Fi1}. In this system, the $R^{target}_{i,j}$  represents the RSS measurements that is transmitted from the $j$-th AP to the $i$-th AP when there is a target within $D$, while $R^{vacant}_{i,j}$ represents the RSS measurements when there is no target. Accordingly, the difference in RSS measurements is denoted by $\Delta \mathrm{R}_{i,j}$ and can be formulated as:
\begin{equation}
\Delta \mathrm{R}_{i, j}=\mathrm{R}_{i, j}^{\text {target }}-\mathrm{R}_{i, j}^{\text {vacant }},
\end{equation}
where
\begin{equation}
\Delta \mathrm{R}=\left[\begin{array}{cccc}
\Delta \mathrm{R}_{1,1} & \Delta \mathrm{R}_{1,2} & \cdots & \Delta \mathrm{R}_{1, D} \\
\Delta \mathrm{R}_{2,1} & \Delta \mathrm{R}_{2,2} & \cdots & \Delta \mathrm{R}_{2, D} \\
\vdots & \vdots & \ddots & \vdots \\
\Delta \mathrm{R}_{D, 1} & \Delta \mathrm{R}_{D, 2} & \cdots & \Delta \mathrm{R}_{D, D}
\end{array}\right].
\end{equation}

In this setting, the transmitted signal can be absorbed and scattered in all directions, and hence, when the target is moving the RSS measurements will be changed accordingly, which produces a different RSS matrix, i.e. each matrix represents a different pattern of RSS measurements. These resulted different patterns can be described as classification problem and could be solved using DL algorithms \cite{Wang2020}. To this end, $D$ is subdivided into $L$ grid cells and each cell is represented as a class and each class will be treated as a potential class; namely, for each gird cell $l \in\{1,\cdots L\}$, we perform $p \in \{1,\cdots,P\}$ trails and vectorizing the dimensional into one dimensional shape as one input data with location information and trait as $V=[v_{11}\cdots v_{lp} \cdots v_{LP}]$.

The overall framework of proposed approach to solve the DFL problem is depicted in Fig. \ref{fig:Figure_2}. First step is to collect the DFL data, and then, divide it into grid cells in order to be fed the deep learning architecture. In this framework, the CDBN used as feature recognizing and extractor in the pre-training stage that will be merged with an autoencoder for fine-tuning. Afterwards, the extracted features are employed for classification through via softmax regression block to determine the potential grid. The RBM is probabilistic joint distribution with bipartite graph with visible input variable $v = [v_1,v_2,\cdots,v_N]^T$ of dimension $D$ and set of hidden random variables $h$ with dimension $K$, $h = [h_1,h_2,\cdots,h_N]^T$, it has symmetrical connections between the visible and hidden layers defined by a weight matrix $W\in R^{(D \times K)}$.
The RBM can be represented as Markov random field, where the visible units represent the input data, and the hidden units are the latent factors. Hence, the weights represent the statistical relationship encoding  between the visible nodes and hidden nodes. In term of Bernoulli distribution, the formal probabilistic of RBM is defined by its energy function as follows: \vspace{-1mm}
\begin{equation}\vspace{-1mm}
    p(v, h)=\frac{1}{Z} \exp (-\mathrm{E}(v, h)),
\end{equation}
where $Z$ represents the normalization constant. The energy function is defined as follows\vspace{-1mm}
\begin{equation}\vspace{-1mm}
\mathrm{E}(v, h)=-\sum_{i=1}^{D} \sum_{j=1}^{K} v_{i} W_{i j} h_{j}-\sum_{j=1}^{K} b_{j} h_{j}-\sum_{i=1}^{D} c_{i} v_{i},
\end{equation}
where $c_i$ account for the visible unit biases, $b_i$ represent the hidden unit biases, and  $W_{i j}$ are the connection weights between the hidden and visible layers. Since the input layer have a real value, the energy function can be defined by adding quadratic term as expressed below: \vspace{-1mm}
\begin{equation}\label{eqn:energy_fun}\vspace{-1mm}
\mathrm{E}(v, h)=\! \frac{1}{2} \sum_{i=1}^{D} v_{i}^{2}-\!\! \sum_{i=1}^{D} \sum_{j=1}^{K} v_{i} W_{i j} h_{j}- \!\! \sum_{j=1}^{K} b_{j} h_{j}-\!\! \sum_{i=1}^{D} c_{i} v_{i}.
\end{equation}

The energy function in (\ref{eqn:energy_fun}) represents the conditional probability distribution and joint probability distribution, which leads to define the hidden unit as conditionally independent with visible layer. The binary hidden layers are conditionally independent on visible layers and can be found by using Bernoulli random variables as follows: \vspace{-2mm}
\begin{equation}\vspace{-3mm}
p\left(h_{j}=1 \mid v\right)=\sigma\left(\sum_{i} W_{i j} v_{i}+b_{j}\right),
\end{equation}\vspace{-1mm}

\noindent 
where  $\sigma$  stands for sigmoid function. Likewise, if the visible layer is conditionally independent on hidden layer and binary valued the Bernoulli random variable becomes: 
\begin{equation}\label{eqn:BRV}
p\left(v_{i}=1 \mid h\right)=\sigma\left(\sum_{j} W_{i j} h_{i}+c_{i}\right).
\end{equation}

\begin{figure}[t]\vspace{-3mm}
	\includegraphics[width=0.5\textwidth]{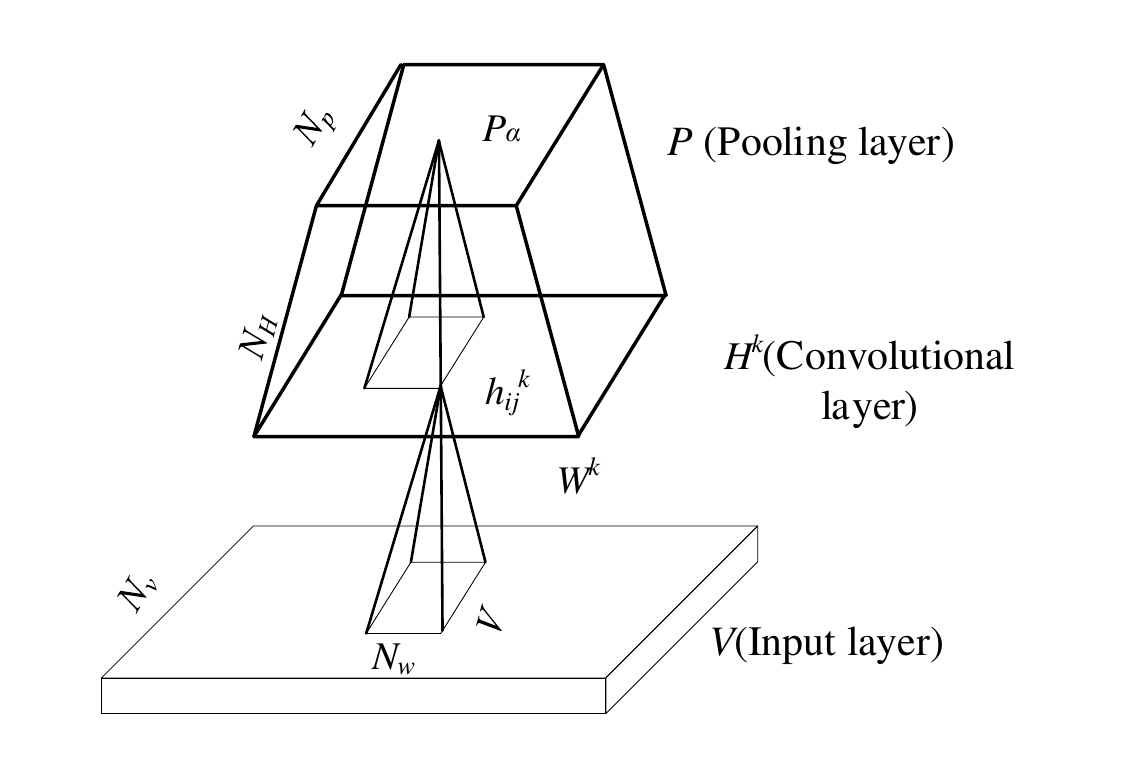} \vspace{-10mm}
	\caption{Construction of a CRBM with probabilistic max-pooling. The CDBN is a hierarchical generative model composed of stacked CRBMs.} \label{fig:CRBM_architecture}\vspace{-5mm}
\end{figure}

\subsection{Convolutional Restricted Boltzmann Machine (CRBM)}
The CRBM consists of RBMs integrated with a convolutional neural network (CNN) that have convolutional and pooling layers. In CRBM, the weights of all locations are shared by the convolutional layer. Meanwhile, the translation-invariant representations are extracted by pooling layer via aggregating the features resulted from the convolutional layer. For probabilistic maximum pooling, the energy function takes the form in (\ref{eqn:BRV}), which allows the bidirectional inference \cite{Lee2011}, which can be defined as:\vspace{-2mm}
\begin{eqnarray}\label{eqn:modified_ef}
E(v, h)=-\sum_{k=1}^{K} \sum_{i, j=1}^{N_{H}} \sum_{r, s=1}^{N_{w}} h_{i j}^{k} W_{r s}^{k} v_{i+r-1, j+s-1} \nonumber \\ 
-\sum_{k=1}^{K} b_{k} \sum_{i, j=1}^{N_{H}} h_{i j}^{k}-c \sum_{i, j=1}^{N_{v}} v.
\end{eqnarray}

In (\ref{eqn:modified_ef}), $v$ denotes the input dataset, $c$  and  $b_k$ account for the shared weights of the location $(i,j)$ over all the hidden layers within the same group.
Each CRBM consists of a visible layer ($V$) that consist of $N_v \times N_v$ array, a hidden layer ($H$) that consist of $N_H \times N_H$ array, and a pooling layer as presented in Fig. \ref{fig:CRBM_architecture}. The CRBM has $K$ convolutional kernels of size $N_w \times N_w$ array such that $N_{W} \stackrel{\Delta}{=} N_{V}-N_{H}+1$ resulting in hidden unit. In this context,
the Gibbs sampler can be applied to approximate the conditional distribution by a discrete set of points, similar to the standard RBM, as detailed below. \vspace{-2mm}


\begin{small}
\begin{subequations}
\begin{eqnarray}
 && \hspace{-12mm} E(v, h)=-\sum_{k=1}^{K} h^{k} \!\! \cdot\left(\tilde{W}^{k} * v\right)\!-\!\!\sum_{k=1}^{K} b_{k} \sum_{i, j} h_{i, j}^{k} \!-c \! \sum_{i, j} \! v_{i j},\!\!\! \label{eq9}\\
&& \hspace{-12mm}  P\left(h_{i j}^{k}=1 \mid v\right)=\sigma\left(\left(\tilde{W}^{k}{ }_{*} v\right)_{i j}+b_{k}\right), \label{eq10}\\
&& \hspace{-12mm} P\left(h_{i j}=1 \mid h\right)=\sigma\left(\left(\sum_{k} W^{k} * h^{k}\right)_{\!\!i j}+c\right). \label{eq11}
\end{eqnarray}
\end{subequations}
\end{small}




\subsection{Convolutional deep belief network (CDBN)}
The CDBN is a multi-layer model extracts high-level features from the low-level ones using a hierarchical structure: lower layers extract low-level features and feed into the higher layers. In other words, the CDBNs are hierarchical generative models composed of multiple stacked CRBMs. This architecture is based on a novel concept called  probabilistic max-pooling.
In general, convolutional networks frequently require two kinds of alternating layers: ``\textit{convolutional}'' and ``\textit{pooling}'' layers, that will shrink the performance of the detection layer by a constant factor, where each unit in pooling layer estimates the maximum value of the detection layer in small region and that will reduce the computational complexity.
In our implementation, the detection and pooling layers consist of $K$ groups, the pooling layer partitioned the detection layer into two or three blocks, where each block ($\alpha$) is connected to the same pooling layer $P_{\alpha}^{k}$. In other words,  the sampling of the visible layer $V$ produces the hidden layer $H$, which takes place in a parallel manner as a mulitmodal distribution by sampling the $h^k_{ij}$ for each block $\alpha$. Next, each block will be sampled independently as a multimodal function,  where  $h^k_{ij}$ is contained as block $\alpha$, i.e., $(i,j) \in B$, and that leads to increase the energy by turning on unit  $h^k_{ij}$, and the conditional probabilities are 

\begin{small} \vspace{-3mm}
\begin{subequations}
\begin{eqnarray}
 && \hspace{-12mm} P\left(h_{i, j}^{k} \!=\! 1 \mid v, h^{\prime}\right) \!=\! \frac{\exp \left(I\left(h_{i, j}^{k}\right)+I\left(p_{\alpha}^{k}\right)\right)}{1+\sum\limits_{\left(i^{\prime}, j^{\prime}\right) \in B_{\alpha}} \!\!\!\!\!\! \exp \left(I\left(h_{i, j}^{k}\right)+I\left(p_{\alpha}^{k}\right)\right)}, \label{eq12}\\
 && \hspace{-12mm} P\left(p_{\alpha}^{k}\!=\!0 \mid v, h^{\prime}\right) \!=\! \frac{1}{1+\sum\limits_{\left(i^{\prime}, j\right) \in B_{\alpha}^{\prime}} \exp \left(I\left(h_{i, j}^{k}\right)+I\left(p_{\alpha}^{k}\right)\right)}. \label{eq13}
\end{eqnarray}
\end{subequations}
\end{small}

As an alternative way to represent the block Gibbs sampling, which will be used  to estimate the posterior distribution. 
In our model, the overcomplete of the data representation  is much larger than the input. Since the first hidden layer contains roughly $K$ groups of units, models with unnecessary or overly detailed features risk learning trivial solutions. A frequently implemented solution is to require a sparse to activate only a tiny fraction of the available units. To encourage each hidden unit group to have a mean activation near a small constant, we regularize the objective function (log-likelihood). Thus, we have determined the following update, followed by contrastive divergence update, appears to be an effective strategy in the field: \vspace{-3mm}

\begin{equation}
\Delta b_{k}^{\text {sparsity }} \propto p-\frac{1}{N_{H}^{2}} \sum\nolimits_{i, j} P\left(h_{i, j}^{k}=1 \mid v\right), \vspace{-1mm}
\end{equation}
where $p$ represents the target sparsity, and each input data is treated as a mini-batch. The learning rate for sparsity update is chosen to make the hidden group’s average activation close to the target sparsity. 
Since this gradient required a high computational cost, a contrastive-divergence learning is used to address this issue, \cite{Hinton2002} proved that the approximate the log-likelihood gradient is sufficient as a replacement to calculate the second term of log-likelihood gradient that needs only a few iterations that modeled by using Gibbs sampling distribution, based on the mentioned theory, the key parameter is derived as follows: \vspace{-3mm}

\begin{small}
\begin{subequations}
\begin{eqnarray}
 && \hspace{-12mm} \Delta W^{k} \propto \frac{1}{N_{H}^{2}}\left(\tilde{Q}^{(0), k} * V^{(0)}-\tilde{Q}^{(n), k} * V^{(n)}\right), \label{eq15}\\
 && \hspace{-12mm}  \Delta b_{k} \propto \frac{1}{N_{H}^{2}} \sum_{i j}\left(Q_{i j}^{(0), k}-Q_{i j}^{(n), k}\right)+\Delta b_{k}^{\text {sparsity }},  \label{eq16}\\
 && \hspace{-12mm}  \Delta c \propto \frac{1}{N_{v}^{2}} \sum_{i j}\left(V_{i j}^{(0)}-V_{i j}^{(n)}\right). \label{eq17}
\end{eqnarray}
\end{subequations} \vspace{-4mm}
\end{small}



\noindent
where $Q^{(0)}$ represents the posterior computation for (\ref{eq12}) and (\ref{eq13}), $k$ represents the number of hidden layers, by updating rules of (\ref{eq15}), (\ref{eq16}), and (\ref{eq17}), CDBN block can be trained efficiently, the unsupervised pre-training process presented in \textbf{Algorithm 1}. 

\begin{algorithm}\label{alg1}
	\SetAlgoLined
	\SetKwInOut{Input}{Input}
	\Input{$V$, epoch number, RBM number $N$}
	\SetKwInOut{Output}{Output}
	\Output{$\nabla_o w_{ij}, \nabla_o c_j$}
	\BlankLine
	Initialize $\nabla w_{ij}, \nabla b_i, \nabla c_j$ randomly, $\mathbf{v}_{dm}^1\leftarrow \mathbf{v}$\\
	Calculate the posterior $Q^{(0)}$ using (\ref{eq12}) and (\ref{eq13}) \\
	Sample $H^{(0)}$ from $Q^{(0)}$ \\
	\While{for each epoch}{
	\While{$ n \leq N$}
	{Sample $V^n$ from $P(v|H^{(n-1)})$ in (\ref{eq10})\\
	 Calculate the posterior  $Q^{(n)}=P(H|v^{n})$ using  (\ref{eq12}) and (\ref{eq13})\\
	 Sample $H^{(0)}$ from $Q^{(0)}$ \\
	}
	Update biases and weight with sparsity regularization and contrastive divergence using (\ref{eq15}), (\ref{eq16}), and (\ref{eq17}), respectively.\\
	Repeat until convergence\\
	\textbf{Unsupervised of autoencoder}\\
	Initialize: $\Delta W^{k}$, $\Delta b_{k}$, and $\Delta c$  from (\ref{eq15}), (\ref{eq16}), and (\ref{eq17}), respectively.\\
	\For{$n=1:N$}{
	Compute the decoder output using (\ref{eqn:eq19})
	}
	Compute MSE using (\ref{eqn:eq20})\\
	Fine-turning via backpropagation until convergence
	}
	\caption{The proposed algorithm}
\end{algorithm}

After a series training of CDBN, the data will be unrolled as an autoencoder which constrict of two phases: encoder and decoder. The encoder output of CDBN can be written as 
\begin{equation}\label{eqn:eq18}
a(v)=f\left(c_{j}+\sum_{i} w_{i j} \frac{v_{i}}{\sigma_{i}^{2}}\right).
\end{equation}

The decoder output after unrolling can be reconstructed as 
\begin{equation}\label{eqn:eq19}
d(\boldsymbol{v})=f\left(c_{j}+\sum_{i} \sigma_{i}^{2} w_{i j}^{T} a(\boldsymbol{v})\right).
\end{equation}

Next, fine-tuning for optimal reconstruction of mean square error (MSE) is: \vspace{-3mm}
\begin{equation}\label{eqn:eq20}
\operatorname{MSE}(\boldsymbol{V})=\frac{1}{s} \sum_{i=1}^{s}(d(v)-v)^{2}
\end{equation}
where $s$ represents the input dataset $v$. The purpose of autoencoder is to build the input data that does not have any information of class labels. Afterwards, a softmax regression is applied to estimate the optimal parameters for the output layer to predict the output class by using labeled data. Subsequently, cross entropy error function is used to estimate the prediction error between predicted data  $y^{\prime}$ and the real data $y$, with a cost function that is defined below:\vspace{-2mm}
\begin{equation}\label{eqn:eq21}
J(\theta)=\frac{1}{s} \sum_{i=1}^{s} y_{s} \times \log \left(y_{s}^{\prime}(\theta)\right), \vspace{-1mm}
\end{equation}
where $\theta$ represents the parameters of the encoder network.  

\section{Performance Evaluation}
In this study, a real experimental dataset from the sensing and processing across networks (SPAN) Laboratory at the University of Utah is used \cite{Wilson2010}. The area of interest $D$ in the experiment setup is divided into 36 grid cells and covered by a wireless sensor network consists of 28 wireless sensor nodes (APs). This wireless sensor network operates within 2.4 GHz frequency band based on the IEEE 802.15.4 protocol. Each AP is placed in three feet apart in a square area with heights of three feet above the ground to construct 36 grid cells. In this setup, 30 different RSS measurements were collected and send to the computer withing a short period of time, where the data will be partitioned into training and test samples. The data dimension for the training consists of $784 \times 900$, while the testing data includes $784 \times 1$. Localization accuracy is used as a metric to evaluate the performance of the proposed algorithm. 
Furthermore, in real-world scenarios, noise from the surrounding environment is inevitable, which definitely disturbs the DFL  performance. Therefore, a random Gaussian noise is added to each collected RSS to examine our algorithm robustness against noise and signal distortions. 


The proposed algorithm has many parameters need to be optimized and set, i.e. number of hidden layers, number of epochs and different number of each group in every hidden layer. 
Thus, the number of hidden layers versus classification accuracy is evaluated using the proposed algorithm, and the results are shown in Table \ref{table_1}. Clearly, using thee layers gives the highest accuracy, that is 98.02\%, which will be considered in the rest of the running experiments. 
Additionally, for these three hidden layers, different number of groups are tested to compute the accuracy of CDBN, the highest accuracy is obtained when the number of groups for the first, second and third layers are 28, 36 and 36, respectively.

\begin{table}\centering 
\caption{Classification accuracy versus number of hidden layers}\label{table_1}
\begin{tabular}{|c|c|}
\hline
Number of hidden layers & Classification accuracy (\%) \\ \hline
1                       & 75.13                        \\ \hline
2                       & 89.21                        \\ \hline
3                       & 98.02                        \\ \hline
\end{tabular}\vspace{-5mm}
\end{table}

\begin{figure}[t]
	\includegraphics[width=0.52\textwidth]{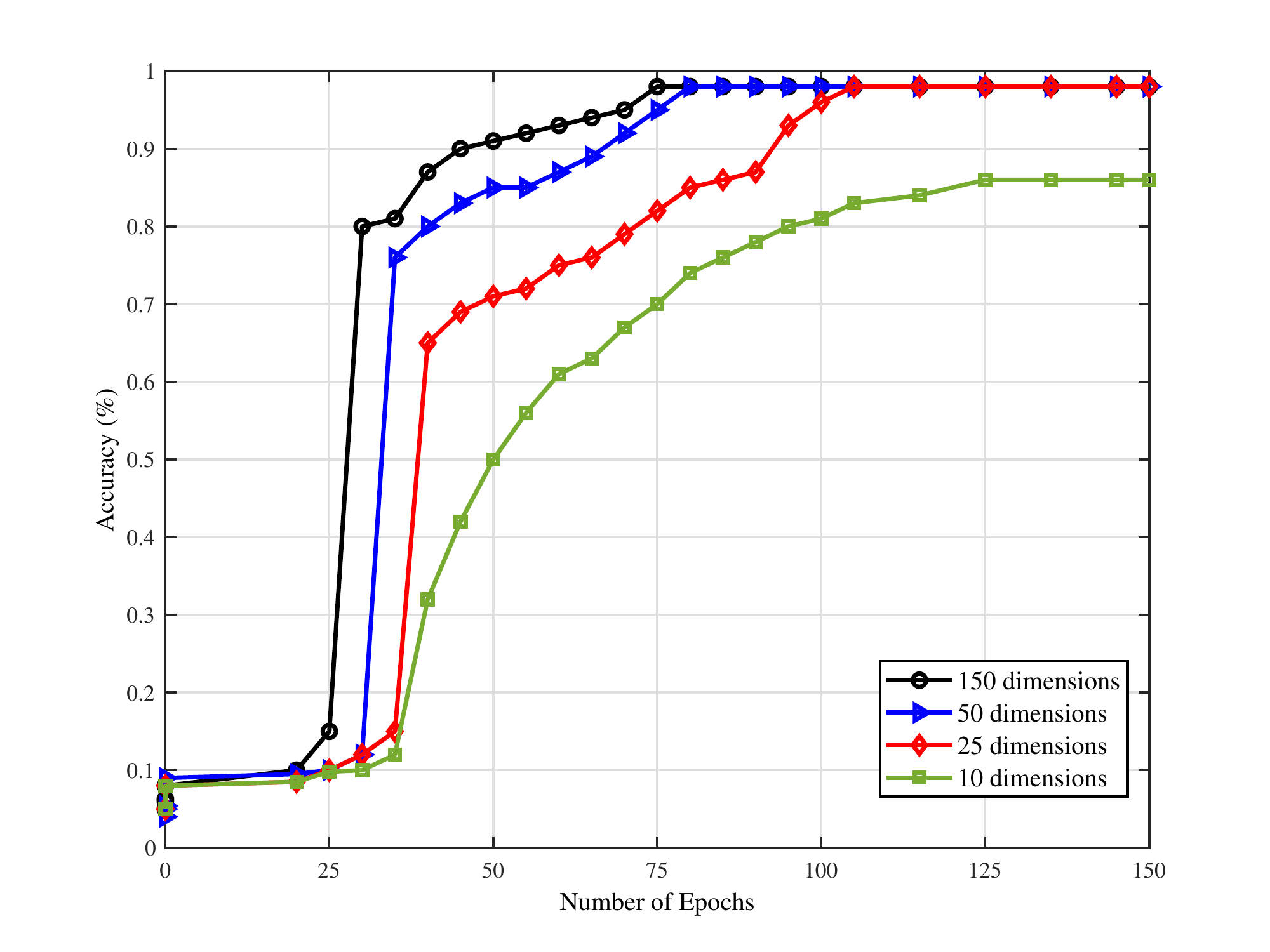}\vspace{-4mm}
	\caption{Localization performance of the CDBN-AE method with data from 10 to 150 dimensions versus the number of epochs.} \label{fig:Fig4}\vspace{-4mm}
\end{figure}

To assess the performance of the proposed algorithm in terms of dimensional reduction and feature extraction, the RSS dimensions are reduced from 784 dimensions to 150, 50, 25, and 15 dimensions and used to evaluate the classification accuracy
as shown in Fig. \ref{fig:Fig4}. Obviously, the highest classification accuracy (98.02\%) is achieved when 150 dimensions with around 35 epochs. Beyond the 35 epochs, the classification accuracy is stable as the epoch number increased. Meanwhile, the 25 dimensions reaches classification accuracy of 97.8\% after a 100 number of epochs. In addition, the 15 data dimensions has the lower classification accuracy with respect to all other numbers despite the increment in the epoch numbers. 
Interestingly, Fig. \ref{fig:Fig4} reveals that the classification accuracy during the training stage can reach 97.8\% for 25 dimensions with a slight increase in the number of epochs. 
These findings validate the outstanding performance of the proposed CDBN-based autoencoder (CDBN-AE) method in performing feature extraction along with data dimension reduction.

\begin{figure}[t]
	\includegraphics[width=0.45\textwidth]{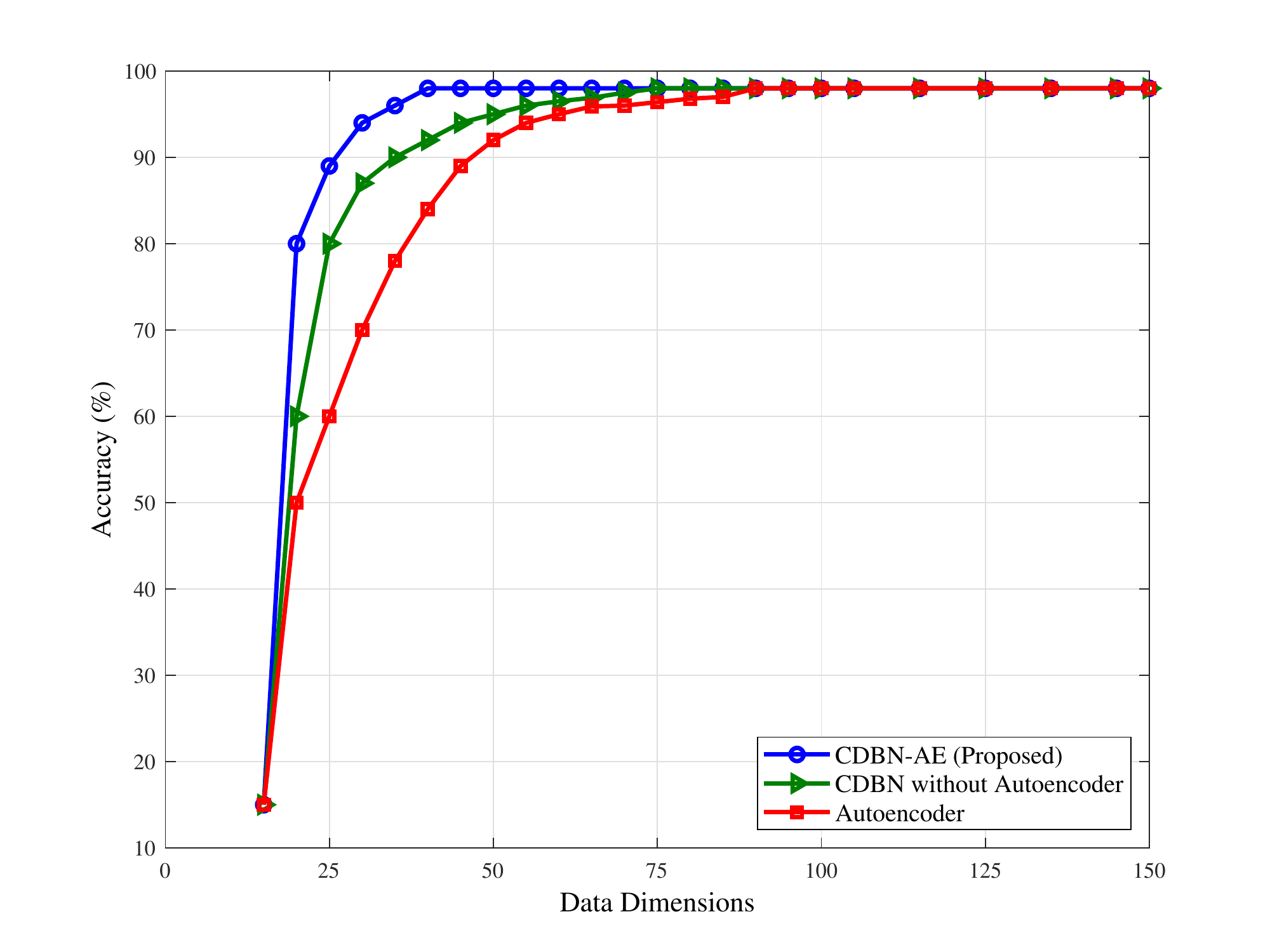}\vspace{-5mm}
	\caption{Localization accuracy performance of the CDBN-AE, CDBN, and autoencoder algorithms versus different dimensional testing data.} \label{fig:Fig5}\vspace{-4mm}
\end{figure}

Intuitively, using large data dimensions during the testing stage increases the computational complexity and slows down the classification process. Thereby, it is critical to find an optimal data dimension that provides an acceptable accuracy. To this end, after completing the DFL training procedure, reduced data ranging from 3 to 150 dimensions is used for the testing stage, and the results are shown in Fig. \ref{fig:Fig5}, which presents the classification accuracy versus testing data dimensions. 
It can be readily seen that the accuracy reaches a maximum of 98.02\% and remains constant when data dimension is greater than 25 for the proposed CDBN-AE method. Further, we compare our CDBN-AE algorithm to the CDBN coder and the autoencoder without CDBN in Fig. \ref{fig:Fig5}, which indicates that the developed CDBN-AE has better localization accuracy with the reduced data dimensions. Thus, the 25 dimensions can be considered as the optimal data dimension for achieving accurate localization.

\begin{figure}[t]
	\includegraphics[width=0.5\textwidth]{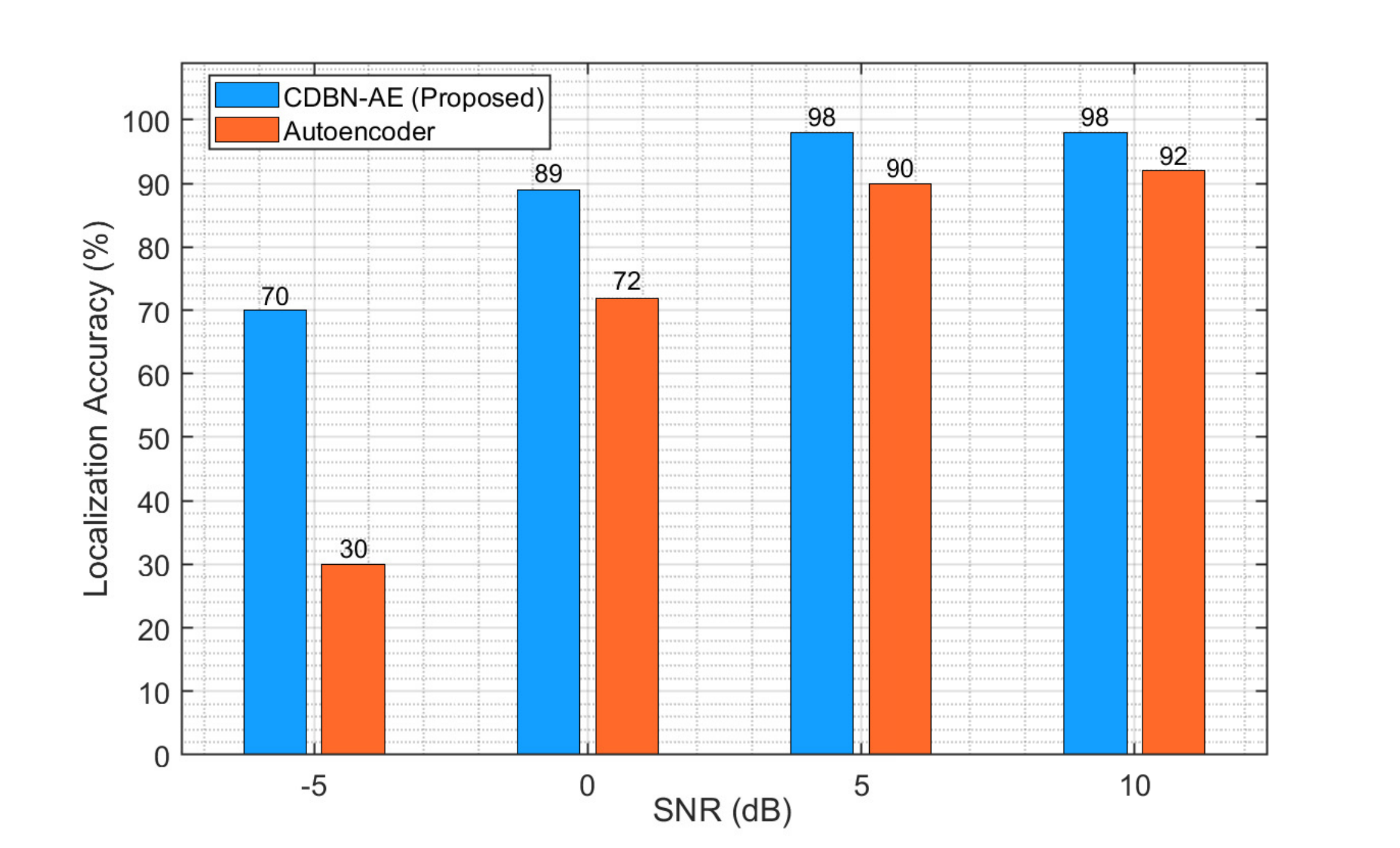}\vspace{-4mm}
	\caption{Comparison of average localization accuracy between the CDBN-AE and autoencoder in the testing stage using noisy dataset with various SNR.} \label{fig:Fig6}\vspace{-5mm}
\end{figure}

Next, the testing performance of the proposed framework is investigated in Fig. \ref{fig:Fig6} using noisy data with varying signal-to-noise ratios (SNRs) and different data dimensions. It is also compared to the autoencoder without CDBN.
Clearly, the localization accuracy of our CDBN-AE method reaches 98\% with 25 data dimensions and when SNR is higher than 10 dB. In addition, the accuracy decreases to 96.2\% and 96\% levels when the SNR values are 0 dB and 5 dB, respectively. Furthermore, at lower SNR values, the location accuracy can be improved to reach 96\% by using the CDBN-AE method and increasing the data dimensions to 150. This demonstrates that the proposed GBRBM-AE algorithm is robust and well-suited  for DFL even at low SNR values.

\section{Conclusion} \vspace{-1mm}
This paper constructed a novel DL-based DFL approach using the variations in RSS measurements in wireless sensor networks. Specifically, the DFL problem is treated as a classification problem having multiple dynamic classes and modeled based on the CDBN with stacked autoencoders. In the proposed framework, pooling layers are added between the convolutional layers of the CDBN, which allow for a reduction in data dimension requirements for localizing a target. The proposed framework offers high levels of location accuracy and robustness against the added random noise from the surrounding environment and the target movements. In particular, the experiment results have showed that with noiseless data the localization accuracy can reach up to 98\% with only 25 data dimensions.
More importantly, considering dataset with noise has proved that the proposed CDBN-AE with 25 data dimensions is able to obtain 98\% accuracy when SNR=10 dB and is resilient to low SNR values around 5 dB with low data dimensions. This proves the applicability of the CDBN-AE approach to real-world DFL scenarios.


\linespread{1}
\vspace{-1mm}
\bibliographystyle{IEEEtran}
\bibliography{IEEEabrv,References}
\end{document}